\documentclass[12pt]{iopart}

\usepackage{graphicx,framed} 
\usepackage{dcolumn} 
\usepackage{bm} 
\usepackage{braket}
\usepackage{xcolor}
\usepackage{amsthm}
\usepackage{hyperref} 
\usepackage[normalem]{ulem} 
\usepackage{mathdots}
\usepackage{amsthm}
\expandafter\let\csname equation*\endcsname\relax
\expandafter\let\csname endequation*\endcsname\relax
\usepackage{amsmath}
\usepackage{amssymb}

\usepackage{comment}

\usepackage[%
style=phys,%
articletitle=false,biblabel=brackets,%
chaptertitle=false,pageranges=false%
]{biblatex} 
\begin{document}

\title[]{Floquet Systems with Continuous Dynamical Symmetries: Characterization, Time-dependent Noether Charge, and Solvability}

\author{Yukio Kaneko$^1$ and Tatsuhiko N. Ikeda$^2$}
\address{$^1$Advanced Research Laboratory, Research Platform, Sony Group Corporation, 1-7-1 Konan Minato-ku, Tokyo, 108-0075 Japan}
\address{$^2$RIKEN Center for Quantum Computing, Wako, Saitama 351-0198, Japan}
\address{$^2$Department of Physics, Boston University, Boston, Massachusetts 02215, USA}
\address{$^2$Institute for Solid State Physics, University of Tokyo, Kashiwa, Chiba 277-8581, Japan}
\ead{$^1$Yukio.A.Kaneko@sony.com and $^2$tatsuhiko.ikeda@riken.jp}

\vspace{10pt}
\begin{indented}
\item[]June 2024
\end{indented}

\begin{abstract}
We study quantum Floquet (periodically-driven) systems having continuous dynamical symmetry (CDS) consisting of a time translation and a unitary transformation on the Hilbert space.
Unlike the discrete ones, the CDS strongly constrains the possible Hamiltonians $H(t)$ and allows us to obtain all the Floquet states by solving a finite-dimensional eigenvalue problem.
Besides, Noether's theorem leads to a time-dependent conservation charge, whose expectation value is time-independent throughout evolution.
We exemplify these consequences of CDS in the seminal Rabi model, an effective model of a nitrogen-vacancy center in diamonds without strain terms, and Heisenberg spin models in rotating fields.
Our results provide a systematic way of solving for Floquet states and explain how they avoid hybridization in quasienergy diagrams. 
\end{abstract}

%
%
%
%
%

\section{Introduction}
Floquet, or periodically driven, systems have attracted renewed attention, partly because recent intense lasers have opened a way to periodically drive quantum systems to control their dynamical properties~\cite{Holthaus2015,Bukov2015,Oka2019,Rudner2020}.
Although the quantum dynamics in such situations are so complicated to simulate theoretically, Floquet theory~\cite{Floquet1883,Shirley1965,sambe1973steady} provides a systematic approximate description, called Floquet states, based on the time-periodic nature of the Schr\"{o}dinger equation.
Nonetheless, obtaining the Floquet states is generically a complicated task because it involves time-dependent problems, even though periodic.

Dynamical symmetries, characteristic of Floquet systems, constrain the properties of Floquet states, helping us to understand the Floquet states without detailed calculations.
These symmetries are combinations of a time translation and a unitary transformation of the physical degrees of freedom and were introduced to explain the selection rules in the high-harmonic generation from graphene and various materials in time-periodic fields~\cite{alon1998selection,alon2002dynamical,alon2004atoms}.
Recently, the dynamical symmetry has been classified systematically based on the group theory~\cite{neufeld2019floquet}, and their consequences have been actively studied~\cite{Engelhardt2021,Wang2021}.

The dynamical symmetry can be extended for the case with an infinitesimal time translation with a unitary transformation \cite{alon2002dynamical}.
In this paper, we refer to this case as continuous dynamic symmetry (CDS).
While there have been discussions on harmonic generation based on CDS \cite{alon2004atoms,pisanty2019conservation,lerner2023multiscale}, a systematic understanding of it has not been achieved.

In this paper, we discuss CDS from theoretical viewpoints and show its three major consequences in time-dependent Schr\"{o}dinger systems: The characterization of Hamiltonians, the existence of the time-dependent Noether charge, and solvability.
The CDS strongly constrains the possible form of time-dependent Hamiltonians [see Eq.(\ref{eq_UHUdagger})], implying the existence of a reference frame in which the Hamiltonian is time-independent.
The time-dependent Noether charge, deriving from Noether's theorem, has a time-independent expectation value, even though the Hamiltonian is no longer a conserved quantity in time-dependent cases.
While obtaining all the Floquet states requires diagonalization of the quasienergy operator of infinite dimension, the CDS reduces it to a finite dimension of the original Hilbert space and explains why there can be degeneracy in quasienergies.

The rest of the paper is organized as follows.
In Sec.~II, we briefly review and summarize our notations for Floquet theory and then introduce the discrete and continuous dynamical symmetries.
In Sec.~III, we derive the three consequences of CDS, which is our main results.
Then, in Sec.~IV, we apply these main results to example models, the Rabi model, an effective model of a nitrogen-vacancy (NV) center in diamonds, and the Heisenberg spin model in a rotating magnetic field.
Section~V summarizes our results and provides concluding remarks.

\section{Dynamical symmetry: Definition and representations}
\subsection{Floquet system on physical and extended Hilbert space}
We begin by briefly reviewing the Floquet theory~\cite{Floquet1883,Shirley1965} and introduce our notations.
Throughout this paper, we consider time-periodic Hamiltonians $H(t)$ of dimension $N$
\begin{equation}\label{eq:Ht_periodic}
H(t+T) = H(t),
\end{equation}
where $T$ denotes the period. 
The Schr\"{o}dinger equation ($\hbar=1$ throughout this paper)
\begin{equation}\label{eq:Schroedinger}
	i\frac{d}{dt}\psi(t) = H(t)\psi(t)
\end{equation}
describes the time evolution of the wave function represented by an $N$-dimensional vector $\psi(t)\in \mathcal{H}= \mathbb{C}^N$.
Here we introduced $\mathcal{H}$ denoting the physical Hilbert space.
Since Eq.~\eqref{eq:Schroedinger} is linear, there exist a set of $N$ solutions $\{\psi_\mu(t)\}_{\mu=1}^N$, and any solution $\psi(t)$ is represented as a linear combination of them.

According to Floquet's theorem, each independent solution $\psi_\mu(t)$ can be written as
\begin{align}
\psi_\mu(t)=e^{-iq_\mu t}u_\mu(t)
\end{align}
with the quasienergy $q_\mu\in\mathbb{R}$ and the periodic Floquet state $u_\mu(t)$ satisfying $u_\mu (t + T) = u_\mu (t) $.
We note that the orthonormality relation
\begin{align}
    \langle\langle u_\mu, u_\nu\rangle\rangle \equiv \frac{1}{T} \int_0^T dt u_\mu(t)^\dagger u_\nu(t) = \delta_{\mu\nu}
\end{align}
holds, where $\langle\langle u,v\rangle\rangle$ is an inner product introduced in the space of periodic states $\mathcal{H}\times S^1$.
The set $\{\mathcal{H} \times S^1, \langle\langle, \rangle\rangle\}$ is called the extended Hilbert (or Sambe~\cite{sambe1973steady}) space.
These arguments imply that solving Eq.~\eqref{eq:Schroedinger} is equivalent to finding all the quasienergies and Floquet states.
If the wave function $ \psi (t) = e ^ {-iqt} u (t) $ is a solution to the Schr\"{o}dinger equation (\ref{eq:Schroedinger}), then the periodic function $u (t)$ satisfies
\begin{align}
	\bar{Q} u(t) = q u(t),
\end{align}
where $\bar{Q} = -i \frac{d}{dt} + H (t) $
is a Hermitian acting on the extended Hilbert space and called the quasienergy operator~\cite{Holthaus2015,Eckardt2017}.

Below, we study the dynamical symmetries of $H(t)$ and its consequences. 
In particular, we define the CDS, determine the corresponding Noether charge, and show that the expected value of the Noether charge is a conserved quantity of this system.
Then, we translate the symmetry and the Noether charge into the extended Hilbert space, showing that it commutes with the quasienergy operator.

\subsection{Discrete and continuous dynamical symmetries (CDSs)}\label{sec:DS}

Now we introduce the dynamical symmetries characteristic of Floquet systems~\cite{alon1998selection,neufeld2019floquet,alon2002dynamical,alon2004atoms}.
These symmetries consist of a unitary transformation on the Hilbert space and a time translation.

We begin by reviewing the discrete dynamical symmetry
\begin{align}\label{eq:discreteDS}
    H\left(t+\frac{T}{n}\right) = S^\dag H(t) S,
\end{align}
where $n\ge2$ is an integer and $S$ is a unitary matrix independent of $t$.
Such symmetry brings about extra constraints, in addition to the periodicity~\eqref{eq:Ht_periodic}, for wave functions within the period $T$.
Using Eq.~\eqref{eq:discreteDS} $k$ times repeatedly, we have
\begin{align}
H\left(t+k \frac{T}{n}\right)=(S^\dag)^k H(t) S^k.
\end{align}
Setting $k=n$, we learn that $[H(t),S^n]=0$ for any $t$.
If $\{H(t) \mid 0\le t\le T\}$ spans $M_N(\mathbb{C})$, $S^n$ is proportional to the identity matrix $I$:
$S^n = e^{i\alpha n}I$, where $\alpha\in\mathbb{R}$ since $S^n$ is unitary.
If $\{H(t) \mid 0\le t\le T\}$ has invariant subspaces, $S^n$ is proportional to the identity in each subspace, and $\alpha$ can be different among the subspaces.

A Hermitian matrix $B$ exists and satisfies $S=\exp(iB)$ since $S$ is unitary.
For this matrix, we have
\begin{align}\label{eq:discreteDS_B}
H\left(t+k \frac{T}{n}\right)=e^{-i k B}H(t)e^{i kB},
\end{align}
and the condition $S^n =e^{inB}=e^{i\alpha n}I$ implies that all the eigenvalues of $nB$ are congruent to $\alpha$ modulo $2\pi$.
Here we note that the phase factor $e^{i\alpha n}$ can be absorbed into $S$ by redefining $Se^{-i\alpha}$ as a new $S$, but we do not do so in this paper.
Instead, we use this degree of freedom to impose $\mathrm{tr}B=0$ since it often simplifies $B$.
We give an example system with the discrete dynamical symmetry in \ref{sec_discrete}.

In this paper, we mainly study the continuous dynamical symmetry (CDS), which is a generalization of Eq.~\eqref{eq:discreteDS_B}.
For the discrete DS, $kT/n$ in Eq.~\eqref{eq:discreteDS_B} is a rational-number multiple of $T$, but we regard $kT/n\to \Delta t$ being an arbitrary real number.
Thus, when there exists a time-independent Hermitian matrix $A$ that satisfies
\begin{align}
    \label{eq_OHO=H}
	H(t+\Delta t) = S_{\Delta t}^\dag H(t) S_{\Delta t}=e^{-i A \Delta t}H(t) e^{i A \Delta t}.
\end{align}
for arbitrary $t$ and $\Delta t$, we define $S_{\Delta t}=e^{i A\Delta t}$ as a CDS and $A$ as its generator.
Without loss of generality, we assume that $A$ is traceless since Eq.~\eqref{eq_OHO=H} is invariant under the replacement $A\to A' = A - \mathrm{tr}(A)I$ with $I$ being the identity matrix.

For the wave function $\psi(t)$, the CDS transformation is defined by
\begin{align}
    \label{eq_trans_psi}
    \psi_{\Delta t}'(t) 
    = S_{\Delta t}U(t+\Delta t, t)\psi(t),
\end{align}
where $U(t',t) = \mathcal{T} e^{-i\int_t^{t'}ds H(s)}$ ($\mathcal{T}$ denotes the time-ordering operator). 
We can see that $\psi'_{\Delta t}(t)$ obeys the same Schr\"{o}dinger equation
\begin{align}
    i\frac{d}{dt}\psi'_{\Delta t}(t) = H(t)\psi'_{\Delta t}(t)
\end{align}
even after the CDS transformation \eqref{eq_trans_psi}, if the Hamiltonian satisfies the transformation rule (\ref{eq_OHO=H}) and $S_{ \Delta t} $ is independent of $t$.

\section{Consequences of continuous dynamical symmetries}\label{sec_consequences}
In this section, we derive three consequences of CDS.
Unlike discrete ones, the CDS is so strong to constrain the form of $H(t)$. Besides, it induces a time-dependent Noether charge and even implies solvability, which enables us to obtain all the Floquet states.

\subsection{Characterization of Hamiltonians}\label{sec:characterization}
Here, we ask an inverse problem for CDS: What kind of $H(t)$ is possible if it has a CDS as in Eq.~\eqref{eq_OHO=H} as well as the periodicity $H(t+T)=H(t)$.

First, we reinterpret Eq.~\eqref{eq_OHO=H} by interchanging $t$ and $\Delta t$ and set, for instance, $\Delta t=0$, obtaining
\begin{align}
    \label{eq_UHUdagger}
    H(t)
    = e^{-iAt}H_0 e^{iAt},
\end{align}
where $H_0\equiv H(0)$.
Its differential form is given by 
\begin{align}\label{eq_dH}
    i\frac{d}{dt}H(t) = [A,H(t)].
\end{align}
Equation~\eqref{eq_UHUdagger} means that the time dependence of the Hamiltonian is determined by the unitary transformation $e ^ {-iA t}$ generated by the Hermitian matrix $A$. 
In other words, the possible Hamiltonians are completely characterized as in Eq.~\eqref{eq_UHUdagger} using a single Hermitian matrix $H_0$ when $A$ is given.
This is a strong constraint on $H(t)$ imposed by the presence of a CDS.
In particular, for $N=2$, one can obtain a general Hamiltonian with CDS by solving (\ref{eq_dH}) (see \ref{apdxB}).
From the expression (\ref{eq_UHUdagger}), we see that the trace of the Hamiltonian is time-independent 
$\frac{d}{dt} {\text tr} H(t) = 0 $.
Let us set $H'(t) = H (t) - \frac{ {\text tr} H (t)}{N} 1 $.
If the wavefunction $\psi' (t)$ is the solution to the Schr\"odinger equation with $H '(t)$, the wavefunction $\psi(t) = e^{-i\frac{{\text tr} H (t)}{N}}\psi'(t)$ solves the Schr\"odinger equation for the original Hamiltonian $H(t)$.
Thus, by solving the Schr\"{o}dinger equation for $H'(t)$, the solution to the original Schr\"{o}dinger equation can be easily obtained.
So, for the rest of the paper, unless otherwise noted, we treat traceless Hamiltonians.

Second, $H(t)$ is further constrained by the periodicity $H(t+T)=H(t)$. Equation~\eqref{eq_UHUdagger} leads to $[H(t),e^{iAT}]=0$ for any $t$.
Suppose that the set of operators $\{H(t)\mid 0\le t<T\}$ does not have a common invariant subspace. Then, $e^{iAT}$ is proportional to the identity matrix $I$:
\begin{align}\label{eq_ExpA_id}
    e^{iAT} = e^{i\alpha T}I,
\end{align}
where $\alpha\in\mathbb{R}$ since $e^{iAT}$ is unitary.
Although the phase factor $e^{i\alpha T}$ can be absorbed into $A$ by the redefinition $A\to A+\alpha I$, we do not do this but use this degree of freedom to impose $\mathrm{tr}A=0$.
In terms of the eigenvalues $\{A_i\}_{i=1}^N$, Eq.~\eqref{eq_ExpA_id} means
\begin{align}
    \label{A=alpha}
    A_{i} = \alpha + M_{i}\omega \quad (M_i \in \mathbb{Z})
\end{align}
for each $i$.
Here the condition $\mathrm{tr}A=0$ imposes $\alpha$ to satisfy $N\alpha=0~(\mod \omega)$. 
If $\{H(t)\mid 0\le t<T\}$ has common subspaces, one can apply the above argument and obtain different $\alpha$ for each subspace.

In summary, a CDS with $A$ determines $H(t)$ in the form of Eq.~\eqref{eq_UHUdagger} using an $H_0$.
In addition, the periodicity imposes the $A$'s properties Eqs.~\eqref{eq_ExpA_id} and \eqref{A=alpha}.
This is a complete characterization of $H(t)$ having a CDS.
We remark that the eigenvalues of $H(t)$ [Eq.~\eqref{eq_UHUdagger}], and hence the $\det H(t)$, are independent of time and given by those of $H_0$.
These properties are useful necessary conditions in judging if there is a CDS since it is generally challenging to find a concrete transformation rule, as we will see in Sec.~\ref{IVB}.

\subsection{Time-dependent Noether charge}
Now we study another consequence of CDS.
Being continuous, this symmetry induces a Noether charge, according to the Noether theorem.
Interestingly, the charge is time-dependent, reflecting the combination of time translation and unitary transformation in the Hilbert space.

To derive the Noether charge, we consider an infinitesimal transformation (\ref{eq_trans_psi}).
Using
\begin{align}
    S_{\Delta t} &= 1 + i\Delta t A + {\cal O}(\Delta t^2),\\
    U(t+\Delta t,t) &= 1-i\Delta t H(t) + {\cal O}(\Delta t^2),
\end{align}
we have
\begin{align}
    \Delta\psi(t) &\equiv  \psi'(t)-\psi(t)
    = -i\Delta t G (t) \psi (t) + {\cal O}(\Delta t^2),
\end{align}
where
\begin{align}\label{eq_defG}
    G(t) = H(t)-A
\end{align}
is the generator of the transformation, {\it i.e.}, the Noether charge.
Equation~\eqref{eq_defG} reflects the fact that the transformation consists of the time translation and the unitary transformation on the Hilbert space since $H(t)$ and $A$ are generators of these two, respectively.

According to the Noether theorem, the expectation value of $G(t)$ is time-independent:
\begin{align}
    \label{eq_dG(t)}
    \frac{d}{dt}(\psi(t)^\dagger G(t) \psi(t)) = 0.
\end{align}
One can also directly confirm this equation using Eqs.~\eqref{eq:Schroedinger} and \eqref{eq_dH}.
As is well-known, the energy expectation value is not a conserved quantity when the Hamiltonian depends on time.
However, in the presence of CDS, the difference between the expectation values of $H(t)$ and $A$ is a conserved quantity
$G(t)$ that possesses the dimension of energy.

For completeness, we translate the time-dependent Noether charge into the extended space language.
To emphasize that we regard $G$ as an operator on the extended space, we introduce the following notation $\bar{G} = H(t) - A$,
which does not have $t$ on the left-hand side like $\bar{Q}$.
Its action onto a periodic $u\in \mathcal{H} \times S^1$ gives $[H(t)-A]u(t)$, which is also periodic.
Note that $\bar{G}$ is Hermitian in the extended Hilbert space.
One can easily show that $\bar{G}$ commutes with the quasienergy operator
\begin{align}
    \label{eq_[bG,Hf]}
    [\bar{Q},\bar{G}] =  -\left( i\frac{d}{dt} H(t) - [A,H(t)]\right) = 0,
\end{align}
where we used Eq.~\eqref{eq_dH}.
In the extended space representation, the conservation law is represented by the usual commutation relation between the charge and Hamiltonian.

\subsection{ Solvability}\label{sec_integrability}
Here, we show the third consequence of the CDS, that is, solvability; The symmetry enables us to construct all the Floquet states.
In a generic Floquet system without the symmetry, finding all the Floquet states is equivalent to diagonalizing $\bar{Q}$ in extended Hilbert space that is infinite dimension.
However, this is in general a difficult task, and approximate perturbation methods, such as the high-frequency expansions~\cite{blanes2009magnus, eckardt2015high, mikami2016brillouin}, are usually utilized.
In contrast, the CDS, if present, reduces the infinite dimension to a finite dimension $N$, as we will see below.

To diagonalize $\bar{Q}$, we recall the commutation relation~\eqref{eq_[bG,Hf]}, which dictates that $\bar{Q}$ and the Noether charge $\bar{G}$ can be diagonalized simultaneously.
So, let us diagonalize $\bar{G}$, i.e., solve
\begin{align}
    \bar{G}u_\mu = [H(t)-A]u_\mu(t) = g_\mu u_\mu(t),
\end{align}
for each real eigenvalue $g_\mu$ and its corresponding periodic state $u_\mu\in \mathcal{H}\times S^1$.
Using Eq.~\eqref{eq_UHUdagger}, we have
\begin{align}\label{eq_H0A}
    (H_0 -A)v_\mu(t) = g_\mu v_\mu(t),
\end{align}
where $v_\mu(t)\equiv e^{iAt}u_\mu(t)$.
Equation~\eqref{eq_H0A} means that $v_\mu(t)$ is, up to the overall phase factor, an eigenstate of $H_0-A$,
\begin{align}\label{eq_H0Aeigen}
    (H_0-A)\Psi_i = Q_i \Psi_i.
\end{align}
The phase factor needs to satisfy the periodicity $u_\mu(0)=u_\mu(T)$, i.e., $v_\mu(T)=e^{iAT}v_\mu(0)$, so we take, e.g.,
\begin{align}
    v_\mu(t) = v_{j,n}(t) = e^{i(\alpha +n\omega)t}\Psi_j,
\end{align}
where $\mu=(j,n)$, $j=1,2,\dots,N$, and $n\in\mathbb{Z}$.
Here we assumed that $H(t)$ does not have an invariant subspace and Eq.~\eqref{eq_ExpA_id} holds true.
Thus we obtain the Floquet states
\begin{align}\label{eq_umu}
    u_\mu(t) = u_{j,n}(t) = e^{i(\alpha+n\omega)t}e^{-iAt}\Psi_j,
\end{align}
which satisfies $u_\mu(T)=u_\mu(0)$ because of Eq.(\ref{eq_ExpA_id}).
One can check that these states are the eigenstates of $\bar{Q}$:
\begin{align}\label{eq_HFeigen}
    \bar{Q}u_{j,n} = (Q_j + \alpha + n\omega) u_{j,n} = q_{j,n}u_{j,n},
\end{align}
where $q_{j,n}\equiv Q_j +\alpha + n\omega$ are the quasienergies.

The above derivation of all the Floquet states requires the diagonalization of the $N\times N$ matrix $H_0-A$, rather than the infinite-dimensional $\bar{Q}$.
This is a remarkable characteristic of CDS, by which we mean solvability.

Physically, the derivation implies a rotational frame in which the Hamiltonian becomes static.
The time dependence of $H(t)$ in Eq.~\eqref{eq_UHUdagger} suggest a unitary transformation $\psi(t) \to \tilde{\psi}(t)\equiv e^{iAt}\psi(t)$, which obeys
\begin{align}
    i\frac{d}{dt}\tilde{\psi}(t) = (H_0 - A )\tilde{\psi}(t),
\end{align}
meaning that the Hamiltonian $H_0-A$ is independent of time in this frame.
All the independent solutions to this Schr\"{o}dinger equation are given by $e^{-iQ_j t}\Psi_j$, leading to the solutions $e^{-iQ_j t}e^{-iA t}\Psi_j$ in the original frame.
Once we extract their periodic part and quasienergy, we obtain the same results as Eq.~\eqref{eq_umu} and \eqref{eq_HFeigen}.
Note that Ref.~\cite{iadecola2013generalized} also provides a similar discussion on the unitary transformation, focusing on Dirac fermions in graphene coupled to a heat bath in the presence of a rotating Kekul\'{e} mass term.

It is noteworthy that the quasienergies $q_{j,n}$ depend only on the Hermitian matrix $H_0-A$,
so two different Hamiltonians with CDS, $H (t) = e^{-iAt} H_0 e^{iAt}$ and $H'(t) = e^{-iA't} H'_0 e^{iA't}$, have the same quasienergies if $H_0-A=H_0'-A'$.
Using this fact, we can derive the complete set of Hamiltonians with a CDS in two-level quantum systems (see \ref{apdxB} for detail).

\section{Examples}
In this section, we discuss example model Hamiltonians having CDS and apply the general properties derived in Sec.~\ref{sec_consequences}.
We analyze the two-level Rabi model, the three-level effective model for a nitrogen-vacancy (NV) center, and the Heisenberg (XXZ) many-spin model in a unified manner from the symmetry viewpoint.

\subsection{Rabi oscillation}
\label{sec4A}
In this section, we consider the Rabi model~\cite{Rabi1937} as a representative Hamiltonian with CDS.
The Hamiltonian is given by
\begin{align}
    \label{eq_RWhamiltonian}
    H_\mathrm{RW}(t) =
    \left(
    \begin{array}{cc}
        \frac{\omega_0}{2} & b e^{-i\omega t}\\
        b e^{i\omega t} & -\frac{\omega_0}{2}
    \end{array}
    \right).
\end{align}
This Hamiltonian is also known in the context of the rotating-field approximation.
As discussed above, we assumed that the Hamiltonian~\eqref{eq_RWhamiltonian} is traceless and $b$ is real without losing generality.
The Hamiltonian (\ref{eq_RWhamiltonian}) is the same with the two-level quantum system under circularly polarized driving that has been historically explored, and the analytic solution is well studied \cite{autler1955stark,holthaus1994generalized}.


One can easily confirm that $H_\mathrm{RW}(t)$ has the following CDS,
\begin{align}
    &H_\mathrm{RW}(t+\Delta t)
    = S_{\Delta t}^\dag H_\mathrm{RW}(t)S_{\Delta t},\\
    &S_{\Delta t} =
    \text{diag} (e^{i \frac{\omega \Delta t}{2}}, e^{-i \frac{\omega \Delta t}{2}})
\end{align}
Here, the corresponding generator $A$ is
\begin{align}
    A = \frac{\omega}{2}\sigma_3,
\end{align}
where $\sigma_3 = \mathrm{diag}(1,-1)$.
Therefore, the time-dependent Noether charge in this model reads
\begin{align}
G(t)=H(t)-A=
\begin{pmatrix}
    \frac{\omega_0-\omega}{2} & be^{-i\omega t} \\
    b e^{i\omega t} & -\frac{\omega_0-\omega}{2}
\end{pmatrix},
\end{align}
whose expectation value is independent of time.

Now we try to obtain the Floquet states following the general argument in Sec.~\ref{sec_integrability}.
For this purpose, we solve Eq.~\eqref{eq_H0Aeigen} with $H_0-A=H_\mathrm{RW}(0)-A$:
\begin{align}
    \label{eq_eigeneqQ}
    \left(
        \begin{array}{cc}
            \frac{\omega_0-\omega}{2} & b\\
            b & -\frac{\omega_0-\omega}{2}
        \end{array}
    \right)
    \left(
        \begin{array}{c}
            \alpha_1 \\
            \alpha_2
        \end{array}
    \right)
    =
    Q
    \left(
        \begin{array}{c}
            \alpha_1 \\
            \alpha_2
        \end{array}
    \right).
\end{align}
The eigenvalues and eigenvectors are obtained as
\begin{align}
    \label{eq_eigenvalueQ}
    Q_\pm &= \pm \sqrt{\frac{1}{4}(\omega-\omega_0)^2+b^2},\\
    \left(
        \begin{array}{c}
            \alpha_{1,\pm} \\
            \alpha_{2,\pm}
        \end{array}
    \right)
    &=
    \frac{1}{\kappa_\pm}\left(
        \begin{array}{c}
            b \\
            \frac{1}{2}(\omega-\omega_0) +Q_\pm
        \end{array}
    \right),
\end{align}
where $\kappa_\pm$ are normalization constants given by $\kappa_\pm = [2Q_\pm^2+(\omega-\omega_0)Q_\pm]^{-1/2}$. 
Thus, all the quasienergies are given by Eq.~\eqref{eq_HFeigen} as
\begin{align}
    \label{eq_quasienergiesRW}
    q_{\pm,n} = Q_\pm + \frac{1}{2}\omega + n\omega,
\end{align}
where we set $\alpha=\omega/2$ since $e^{iAT}=e^{i\frac{\omega T}{2} \sigma_3} = e^{i\frac{\omega}{2}T} I$
in the present model.
We obtain the corresponding Floquet states as
\begin{align}
    u_{\pm,n}(t)&=e^{i(n+\frac{1}{2})\omega t}e^{-i\frac{\omega}{2}\sigma_3}
    \begin{pmatrix}
        \alpha_{1,\pm} \\ \alpha_{2,\pm}
    \end{pmatrix}\\
    &=e^{i n\omega t}    \begin{pmatrix}
        \alpha_{1,\pm} \\ e^{i\omega t}\alpha_{2,\pm}
    \end{pmatrix}
.\label{eq_solRW}
\end{align}

\begin{figure}
    \begin{center}
    \includegraphics[width=0.48\columnwidth]{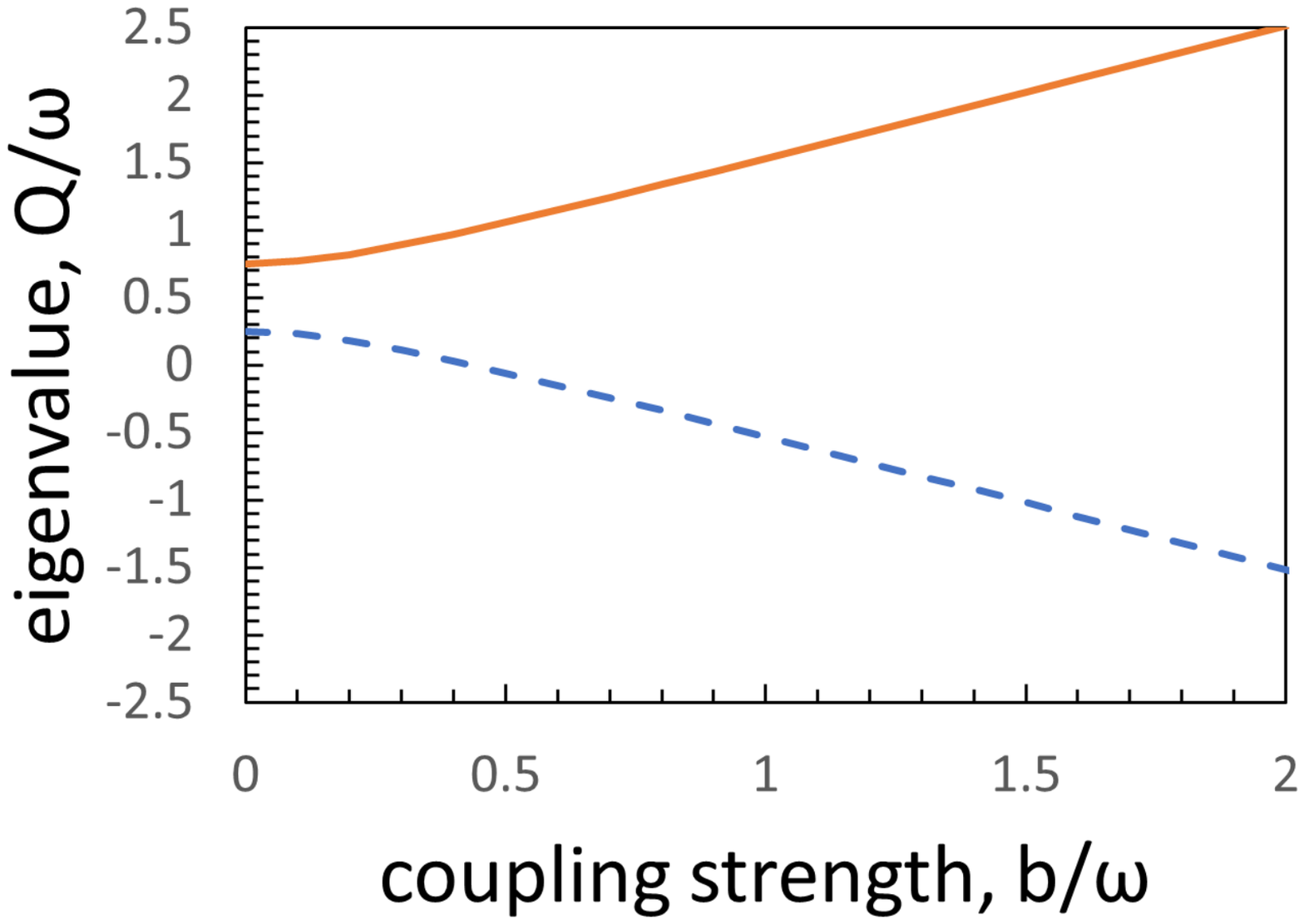}
    \includegraphics[width=0.48\columnwidth]{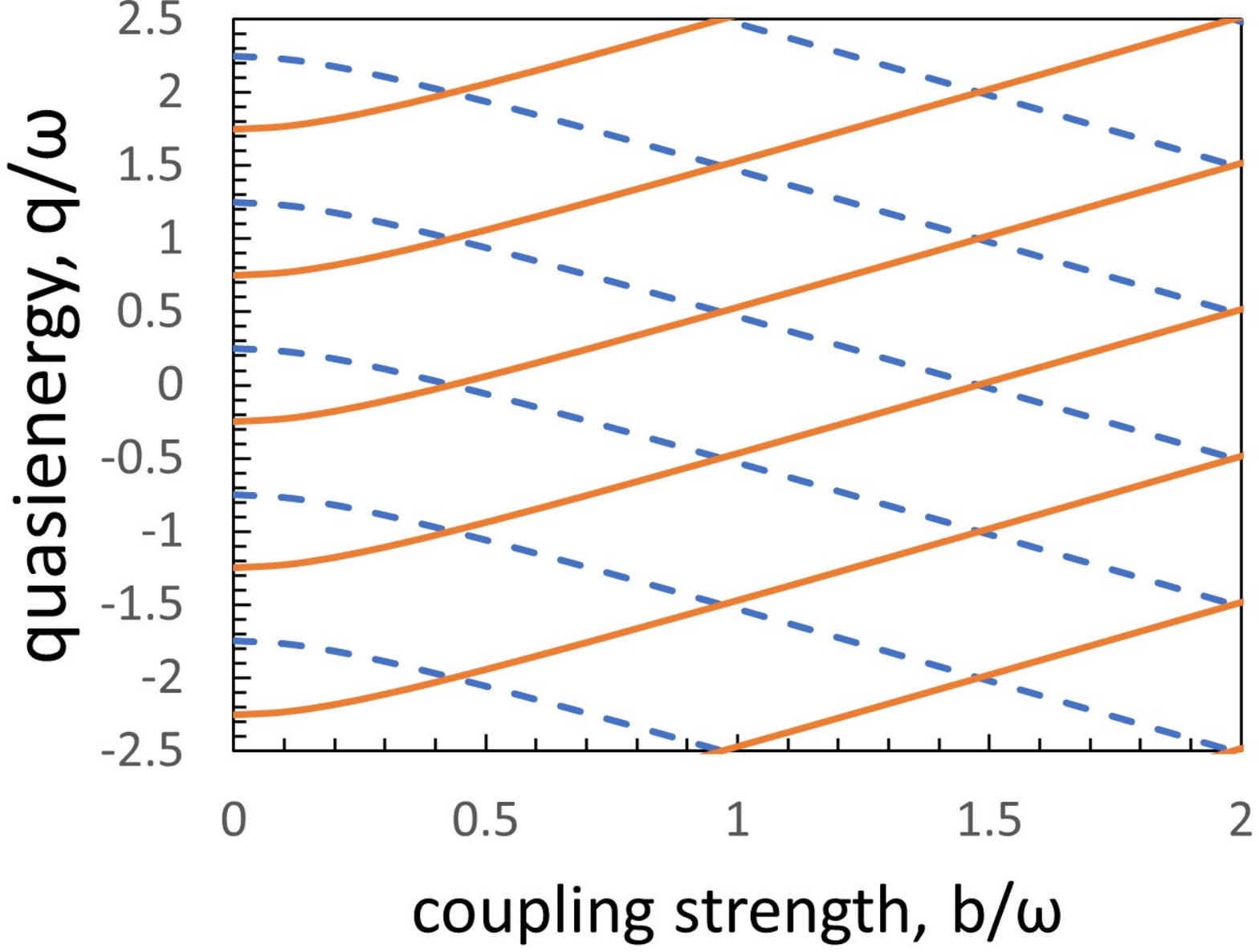}
    \caption{
    Eigenvalues $Q_\pm$ (upper) and quasienergies $q_{\pm,n}$ (lower) for the Rabi oscillation (\ref{eq_RWhamiltonian}) given in \ref{sec4A} with $\omega_0/\omega =0.5$ plotted against coupling strength $b$. Solid (dashed) curves show $Q_+$ ($Q_-$) and coresponding quasienergies for Floquet states approaching $u_{+,n}$ ($u_{-,n}$).}
    \label{figspec1}
    \end{center}
\end{figure}

By virtue of the CDS, the eigenvalue problem of the quasienergy operator is reduced to a two-dimensional eigenvalue problem [Eqs.~(\ref{eq_eigeneqQ}) and (\ref{eq_eigenvalueQ})], and the quasienergies are obtained by their shift by $n\omega$ (\ref{eq_quasienergiesRW}).
The resulting quasienergies have many degeneracies, as shown in Fig.~\ref{figspec1}.
We note that the degenerate pairs have different eigenvalues of the Noether charge $\bar{G}$, which are evaluated as follows:
\begin{align}
    \label{eq_barGrabi}
    \bar{G}u_{\pm,n}=g_{\pm,n}u_{\pm,n}=-\left(n+\frac{1}{2}\right)\omega u_{\pm,n}.
\end{align}
It has been known that the degeneracy is lifted when we add a small counter-rotating term in $H_\mathrm{RW}(t)$ (see, e.g., Appendix A in Ref.~\cite{ikeda2022floquet}).
Our finding here is the symmetry viewpoint of the degeneracy's behavior: The degeneracy is due to the CDS and disappears when the CDS is broken.
For completeness, we show how quasienergies differ when we break the CDS down to a discrete one in \ref{sec_discrete}.

\subsection{Effective model of nitrogen-vacancy center}\label{sec:NV}
\label{IVB}
Let us consider the following effective Hamiltonian of an NV center~\cite{Rondin2014}
\begin{align}
    \label{eq_NV center}
	H_\mathrm{NV}(t) &=-B_s S_z +N_z S_z^2\notag\\
 &\qquad +N_{xy}(S_x^2-S_y^2) +H_\mathrm{ext}(t),
\end{align}
where $S_\alpha$ $(\alpha=1,2,3)$ are the $3\times 3$ spin matrices for spin $S=1$, whose elements are taken, e.g.,
\begin{align}
    &S_x =\frac{1}{\sqrt{2}}\begin{pmatrix}
        0 & 1 & 0\\
        1 & 0 & 1\\
        0 & 1 & 0
    \end{pmatrix},
    \ S_y =\frac{1}{\sqrt{2}}\begin{pmatrix}
        0 & -i & 0\\
        i & 0 & -i\\
        0 & i & 0
    \end{pmatrix},\notag\\
    &S_z =\mathrm{diag}(1,0,-1).
\end{align}
In this subsection, we do not impose $\mathrm{tr}(H_\mathrm{NV}(t))$ to vanish, following the standard convention~\eqref{eq_NV center}.
The term $H_\mathrm{ext}(t)$ denotes the coupling to a circularly-polarized ac magnetic field,
\begin{align}
	H_\mathrm{ext}(t) &= -B_d[S_x \cos (\omega t) + S_y \sin(\omega t)].
\end{align}
Recently, strong couplings to external oscillating fields have been achieved in different settings, and Floquet states have been experimentally detected~\cite{Nishimura2022,Mikawa2023}.
For simplicity, we ignore the effects of dissipation~\cite{ikeda2020general,Ikeda2021}

For the CDS being present, the determinant of the Hamiltonian
\begin{align}
    \det H_{NV}(t)
    = -N_z B_d^2 + N_{xy}B_d^2 \cos(2\omega t)
\end{align}
has to be time-independent, which implies $N_{xy} = 0$.
Under this condition,
\begin{align}
    S_{\Delta t}
    = e^{i\omega \Delta t S_z}
    =\mathrm{diag}(e^{i\omega \Delta t},1,e^{-i\omega \Delta t})
\end{align}
satisfies the transformation rule (\ref{eq_OHO=H}).
Also, we find
\begin{align}
    A_\mathrm{NV} &= \omega S_z,\\
    H_{NV}(t) &=
    e^{-iA_{NV}t}
    H_{\mathrm{NV0}}
    e^{iA_{NV}t}.\\
    H_{NV0}
    &= -B_s S_z +N_z S_z^2 -B_d S_x . 
\end{align}
Thus, we obtain the following time-dependent Noether charge
\begin{align}
    G(t) = H_\mathrm{NV}(t) - \omega S_z.
\end{align}

Now we are obtaining the Floquet states.
We solve Eq.~\eqref{eq_H0Aeigen} with $H_0-A=H_{\mathrm{NV0}}-A_\mathrm{NV}$:
\begin{align}\label{eq:eigenNV}
    \left(
        \begin{array}{ccc}
            N_z-B_s-\omega & -B_d/\sqrt{2} & 0 \\
            -B_d/\sqrt{2} & 0 & -B_d/\sqrt{2} \\
            0 & -B_d/\sqrt{2} & N_z + B_s+\omega
        \end{array}
    \right)
    \left(
        \begin{array}{c}
            \beta_1 \\
            \beta_2 \\
            \beta_3
        \end{array}
    \right)
    =
    Q
    \left(
        \begin{array}{c}
            \beta_1 \\
            \beta_2 \\
            \beta_3
        \end{array}
    \right).
\end{align}
This eigenvalue equation is for a $3\times3$ matrix and hence solvable in principle.
For each of the three pairs of the eigenvalues and eigenvectors, Eq.~\eqref{eq_umu} gives
\begin{align}
    u_{j,n} = e^{in\omega t}\begin{pmatrix}
        e^{-i\omega t}\beta_1^{(j)} \\
        \beta_2^{(j)}\\
        e^{i\omega t}\beta_3^{(j)}
    \end{pmatrix},
\end{align}
where we set $\alpha=0$ since $e^{iA_\mathrm{NV}T}=I$,
and $\bm{\beta}^{(j)}$ ($j=1,2,3$) are the eigenvectors of Eq.~\eqref{eq:eigenNV}.
The corresponding eigenvalues are
\begin{align}\label{eq:qQjNV}
    q_{j,n} = Q_j + n\omega,
\end{align}
where $Q_j$ are eigenvalues of Eq.~\eqref{eq:eigenNV}.

\begin{figure}
    \begin{center}
    \includegraphics[width=0.48\columnwidth]{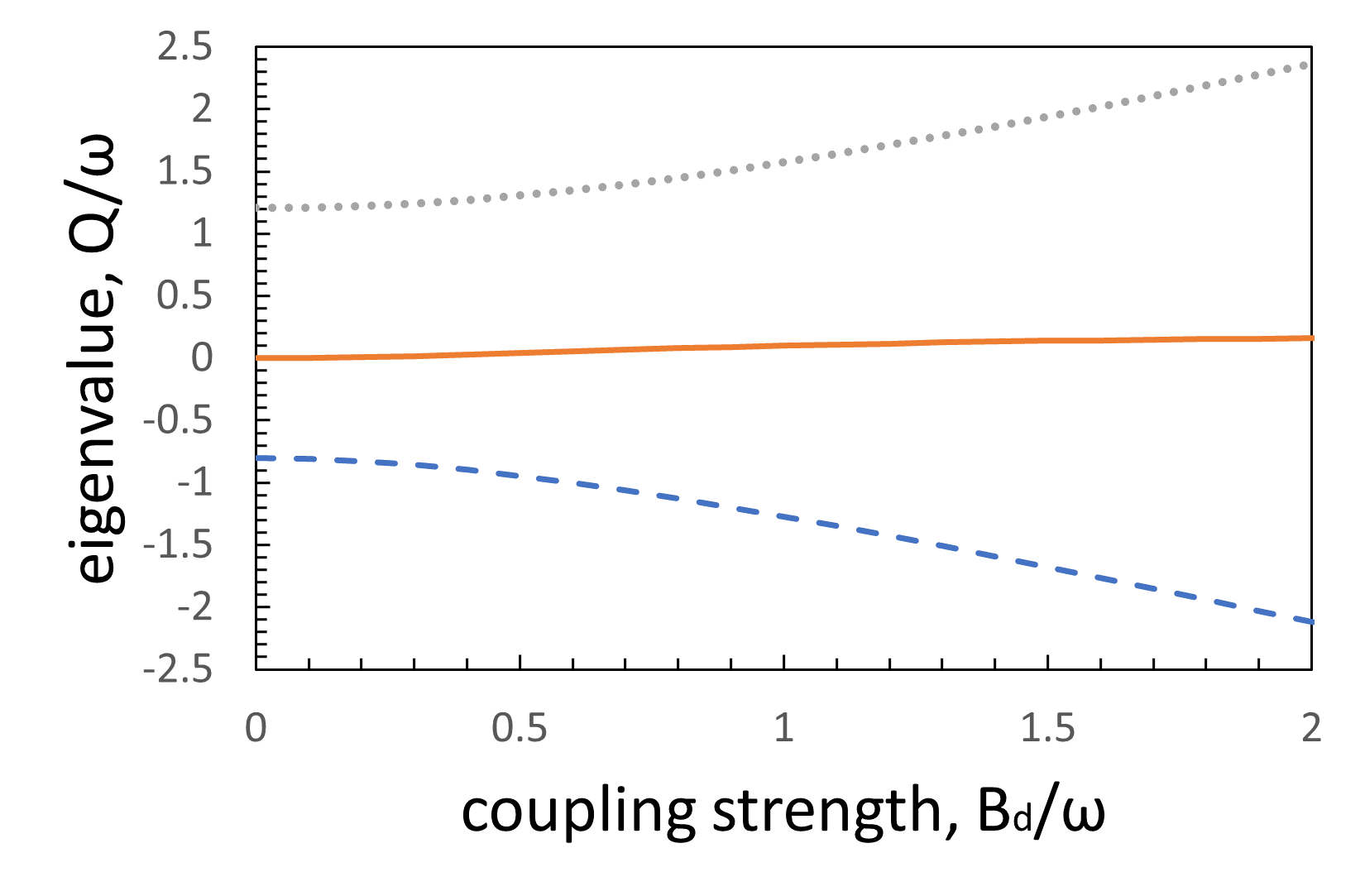}
    \includegraphics[width=0.48\columnwidth]{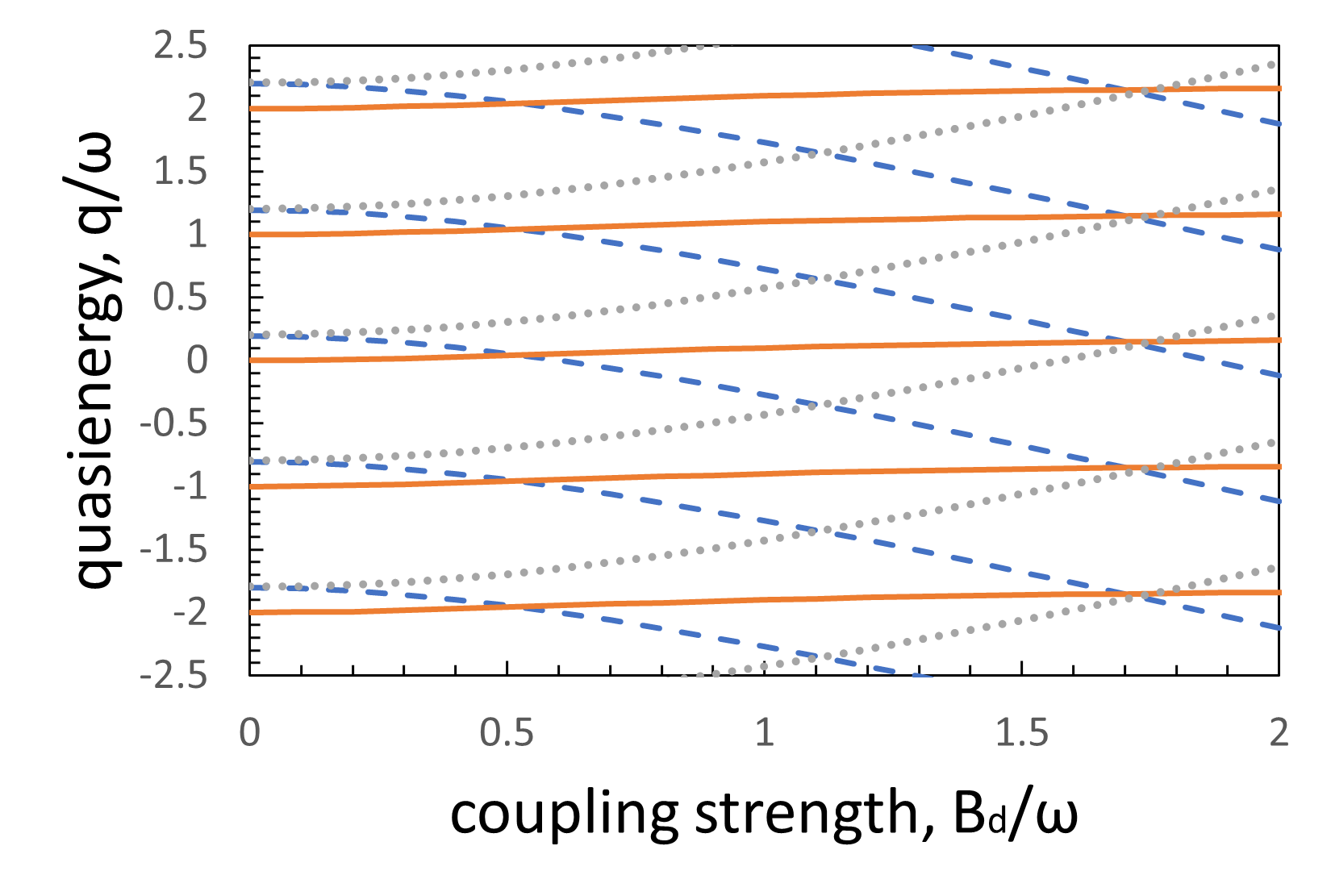}
    \caption{
    Eigenvalues $Q_j$ (upper) and quasienergies $q_{j,n}$ (lower) for the effective model of NV center (\ref{eq_NV center}) given in \ref{sec:NV} with $N_{xy}/\omega =0$, $N_z/\omega = 0.2, B_s/\omega=0.1$ plotted against coupling strength $B_d$.
    }
        \label{figspec2}
    \end{center}
\end{figure}

The above construction of all the quasienergies elucidates why there are many crossings in the quasienergy diagram.
The upper panel of Fig.~\ref{figspec2} shows the three eigenvalues $Q_j$, whereas the lower panel all the $q_{j,n}$.
The quasienergies $q_{j,n}$ are obtained by making replicas for integers $n$ as in Eq.~\eqref{eq:qQjNV}.
Therefore, even if multiple replica quasienergies overlap, no hybridization occurs to create an avoided crossing.
The CDS reduces the eigenvalue problem for $\bar{Q}$ to the finite-dimensional one for $H_0-A$, thereby excluding the avoided crossings.

\subsection{Heisenberg spins in rotating magnetic field}
The argument in Sec.~\ref{sec:NV} can be generalized to a many-spin model.
Here we consider the Heisenberg spin model on a $d$-dimensional lattice in a rotating magnetic field:
\begin{align}
    H(t) &= H_\mathrm{Heis} - B [M^x \cos(\omega t) +M^y \sin(\omega t)],\label{H_spinchain}\\
    H_\mathrm{Heis} &= J\sum_{\langle i,j\rangle} \bm{S}_i\cdot \bm{S}_j,\\
    M^\alpha &= \sum_i S^\alpha_i,
\end{align}
where $S^{\alpha}_i$ $(\alpha=x,y,z)$ denote the spin-1/2 operators acting on site $i$, and $\langle i,j\rangle$ shows each nearest neighbor.

Like in Sec.~\ref{sec:NV}, we have the following CDS,
\begin{align}
    A &= \omega M^z,\\
    H(t) &= e^{-iAt}H_0 e^{iA t},\\
    H_0 &= H_\mathrm{Heis}-B M^x.
\end{align}
These relations lead to the time-dependent charge
\begin{align}
    G(t) = H(t) -\omega M^z,
\end{align}
whose expectation value never depends on time, even if each of $H(t)$ and $\omega M^z$ has time-dependent expectation values.

According to Sec.~\ref{sec_integrability}, once we have the many-body eigenstates
\begin{align}\label{eq:Heiseneigen}
    H_0 \Psi_j = Q_j \Psi_j,
\end{align}
we obtain all the Floquet states
\begin{align}
    u_{j,n} = e^{i n\omega t} e^{-i\omega S^z}\Psi_j
\end{align}
and their quasienergies
\begin{align}
    q_{j,n} = Q_j +n\omega,
\end{align}
where we set $\alpha=0$ since $e^{iAT}=e^{i\omega T M^z}=I$.

We remark that solving Eq.~\eqref{eq:Heiseneigen} is difficult, and it is referred to as nonintegrable for $d\geq2$ in the many-body physics context.
The case of $d=1$ is exceptional, and Eq.~\eqref{eq:Heiseneigen} is Bethe-ansatz solvable.
Irrespective of integrability in these senses, we call our $H(t)$ to be solvable as in Sec.~\ref{sec_integrability} in the sense that the time-dependence of Floquet states is obtained once we have $\Psi_j$.
We also remark that it has been well known that the above unitary transformation eliminates the time dependence of the external field, and we aim here to point out that this fact is understood as a CDS in a unified manner together with other examples in the previous sections.

\begin{figure}
    \begin{center}
    \includegraphics[width=0.48\columnwidth]{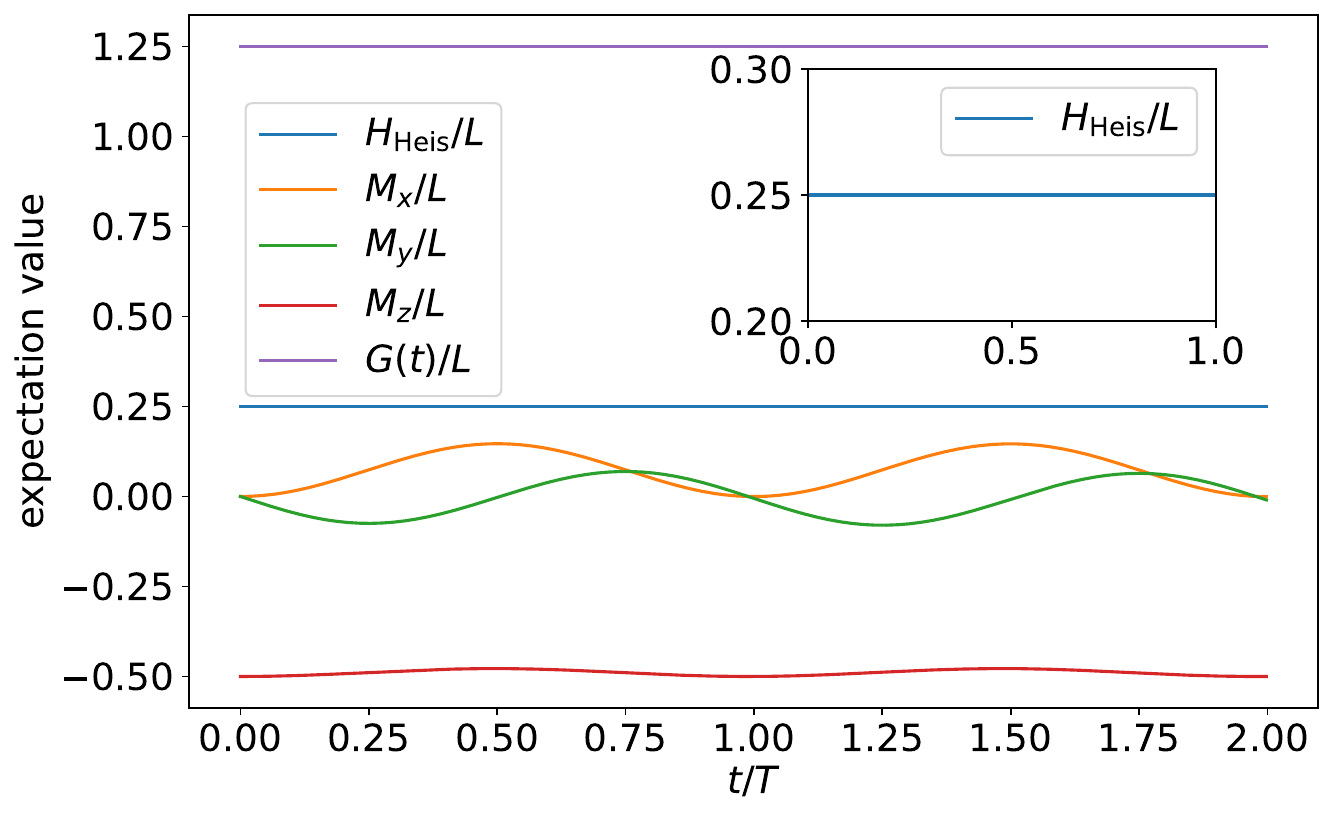}
    \includegraphics[width=0.48\columnwidth]{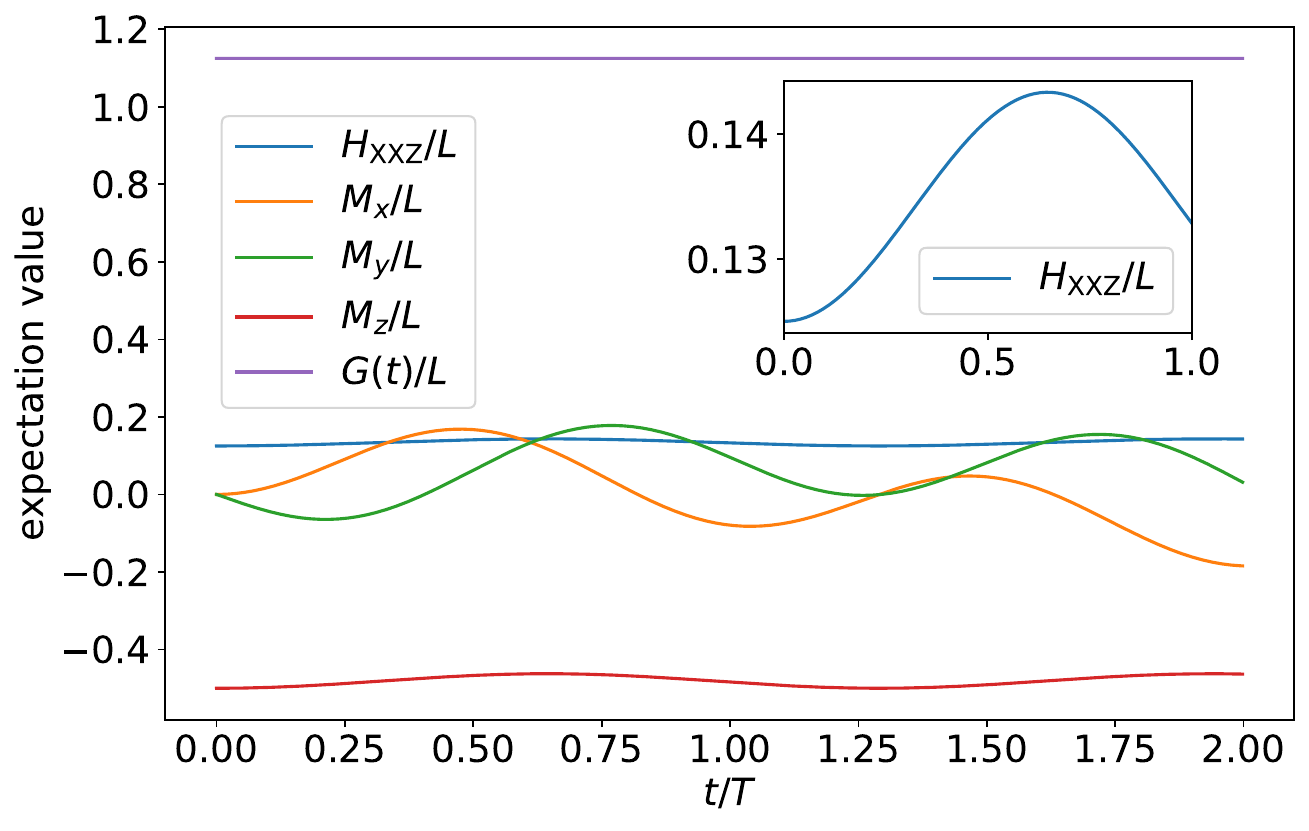}
    \caption{Expectation values of observables (see legends) for the (upper) Heisenberg and (lower) XXZ ($\Delta=1/2$) models.
    The parameters are chosen as $J=1, B=0.3, \omega=2$, and $L=10$.
    The insets are zoomins for the observables $H_\mathrm{Heis}$ and $H_\mathrm{XXZ}$.
    }
    \label{fig:spinchain}
    \end{center}
\end{figure}

Let us explore the physical significance of the Noether charge of CDS.
To visualize their role, we numerically solve the time-dependent Schr\"{o}dinger equation, with the initial state being the all-down state.
We set the parameters as $J=1, B=0.3, \omega=2$, and we consider a $d=1$ chain of length $L=10$ and impose the periodic boundary conditions.
The upper panel of Fig.~\ref{fig:spinchain} shows the expectation values of relevant observables.
Although $M_x,M_y,M_z$ have time-dependent expectation values, the Noether charge as their linear combination
\begin{align}
\label{GHeis}
G(t)=H_\mathrm{Heis} - B [M^x \cos(\omega t) +M^y \sin(\omega t)] -\omega M^z
\end{align}
is constant, as expected from the analytical arguments.
Note that the expectation value of the Hamiltonian $H(t)=G(t)-\omega M^z$ depends on time like in general time-dependent Hamiltonian systems.
The existence of $G(t)$, whose expectation value is time-independent, is a special property of the Floquet system having the CDS.
Note that the undriven Hamiltonian $H_\mathrm{Heis}$ also has a time-independent expectation value, and this is a specialty of the Heisenberg Hamiltonian, for which $[H_\mathrm{Heis},H(t)]=0$ holds for any $t$.
We will discuss its consequences below.

These arguments can basically be generalized to the XXZ model where $H_\mathrm{Heis}=J\sum_i \bm{S}_i\cdot \bm{S}_{i+1}$ is replaced by $H_\mathrm{XXZ}=J \sum_i (S_i^xS_{i+1}^x+S_i^yS_{i+1}^y+\Delta S_i^zS_{i+1}^z)$ with $\Delta \neq1$.
The lower panel of Fig.~\ref{fig:spinchain} shows a similar simulation result for $\Delta=1/2$.
Like in the Heisenberg case ($\Delta=1$), the time-dependent Noether change
\begin{align}
    G(t)=H_\mathrm{XXZ} - B [M^x \cos(\omega t) +M^y \sin(\omega t)] -\omega M^z
\end{align}
has a time-independent expectation value, although each of its components has time-dependent expectation values.
However, the underiven Hamiltonian $H_\mathrm{XXZ}$ is not a conserved quantity, unlike $H_\mathrm{Heis}$ in the Heisenberg case.
This is generic in Floquet systems having CDSs and holds for the Rabi and NV-center models that we discussed above.

Finally, we discuss special property of the Heisenberg spin model.
The Noether charge is explicitly given in Eq. (\ref{GHeis}).
Meanwhile, the commutation relations $[H_\mathrm{Heis},M^i]=0$ ($i=x,y,z$) imply $[H(t),H_\mathrm{Heis}]=0$ ($\forall t$), meaning that $H_\mathrm{Heis}$ is a conserved quantity in this system.
As a result, their difference
\begin{align}\label{eq_E(t)}
E(t) &= H_\mathrm{Heis}-G(t)\\
&= B [M^x \cos(\omega t) +M^y \sin(\omega t)] +\omega M^z
\end{align}
is also conserved.
Thus, we obtain
\begin{align}
\frac{d}{dt}\langle M^z \rangle = -\frac{B}{\omega}\frac{d}{dt}(\langle M^x \rangle \cos(\omega t) + \langle M^y \rangle \sin(\omega t)).\label{eq_ddt_Mz}
\end{align}
The Zeeman energy is generally given by $E = -\bm{B}\cdot \bm{M}$, the Eq. (\ref{eq_ddt_Mz}) can be interpreted as indicating that the time evolution of magnetization of this spin chain arises from the Zeeman energy caused by a rotating magnetic field. 
In summary, it is possible to identify conserved physical quantities through the conservation law of the Noether charge of CDS, which governs the time evolution of operators in the physical frame, even without conducting a detailed analysis of the Hamiltonian.

\section{Discussions and Conclusions}
We have systematically analyzed the continuous dynamical symmetry (CDS) and its consequences in quantum Floquet systems.
Unlike the discrete ones, the CDS is so strong to determine the time dependence of Hamiltonians as in Eq.~\eqref{eq_UHUdagger}.
This special form of time dependence implies the existence of a rotational frame in which the Hamiltonian looks static.
Such frames have been known in several concrete models, but we have provided a unified understanding in terms of the CDS.
The CDS allows us to obtain all the Floquet states out of a finite-dimensional eigenvalue problem~\eqref{eq_H0Aeigen}, uncovering the underlying mechanism for level crossing, instead of hybridization, in the quasienergy diagram in, e.g., the Rabi model.
In the original frame, the CDS manifests as a time-dependent Noether charge~\eqref{eq_defG}, whose expectation value is constant under the time-dependent Schr\"{o}dinger equation.

We leave some extensions of our work for future study.
First, finding more examples of the CDS would be important.
Although our three examples were (effective) spin models, the charge degrees of freedom of electrons may have CDSs.
For example, an electron gas in a $d(\ge2)$-dimensional space under a circularly-polarized (about, e.g., the $z$-axis) ac electric field has a CDS consisting of the spatial rotation.
In this case, the generator $A$ corresponds to the orbital angular momentum $L_z$, and the time-dependent Noether charge is given by $G(t)=H(t)-\omega L_z$.
Second, extensions to non-Floquet systems are of interest.
Without time-periodicity, the Hamiltonian can be more generic, but the time-dependent Noether charge could still be defined.
One possible direction is the study of quasi-periodic systems described in \cite{Zhao2019,pizzi2019period}, where a periodic framework is achieved through time-dependent unitary transformations.
There are also infinite-dimensional examples of non-Floquet systems, such as the system of driven harmonic oscillators \cite{husimi1953miscellanea,dittrich1998quantum}.
Not limited to these two directions, it would be useful to analyze nonequilibrium phenomena from symmetry viewpoints.

\section*{Acknowledgements}
The authors thank E. Igarashi for her contribution in the early stage of the work
and M. Holthaus, Y. Hidaka, and H. Taya for fruitful discussions.
T. N. I. was supported by JST PRESTO Grant No. JPMJPR2112 and by JSPS KAKENHI Grant No. JP21K13852.

\appendix

\section{Example of quantum system with discrete dynamical symmetry}
\label{sec_discrete}
In this section, we consider the following Hamiltonian as an example of a system with discrete dynamic symmetry:
\begin{align}
    H_C(t)
    =
    \left(
        \begin{array}{cc}
            \frac{\omega_0}{2} & b \cos \omega t\\
            b^* \cos \omega t & -\frac{\omega_0}{2}
        \end{array}
    \right).\label{cosdriven}
\end{align}
This system is explored in detail in Ref.~\cite{Shirley1965}.
The Hamiltonian $H_C$ has the following discrete dynamical symmetry:
\begin{align}
    P_{T/2}(t)\psi(t) =& O_{T/2}U(t+T/2, t)\psi(t),\\
	O_{T/2}H(t+T/2) O_{T/2}^{\dagger} =& H(t),\\
    O_{T/2}
    =&
    \left(
        \begin{array}{cc}
            1 & 0\\
            0 & -1
        \end{array}
    \right).
\end{align}

\begin{figure}
    \begin{center}
    \includegraphics[width=0.6\columnwidth]{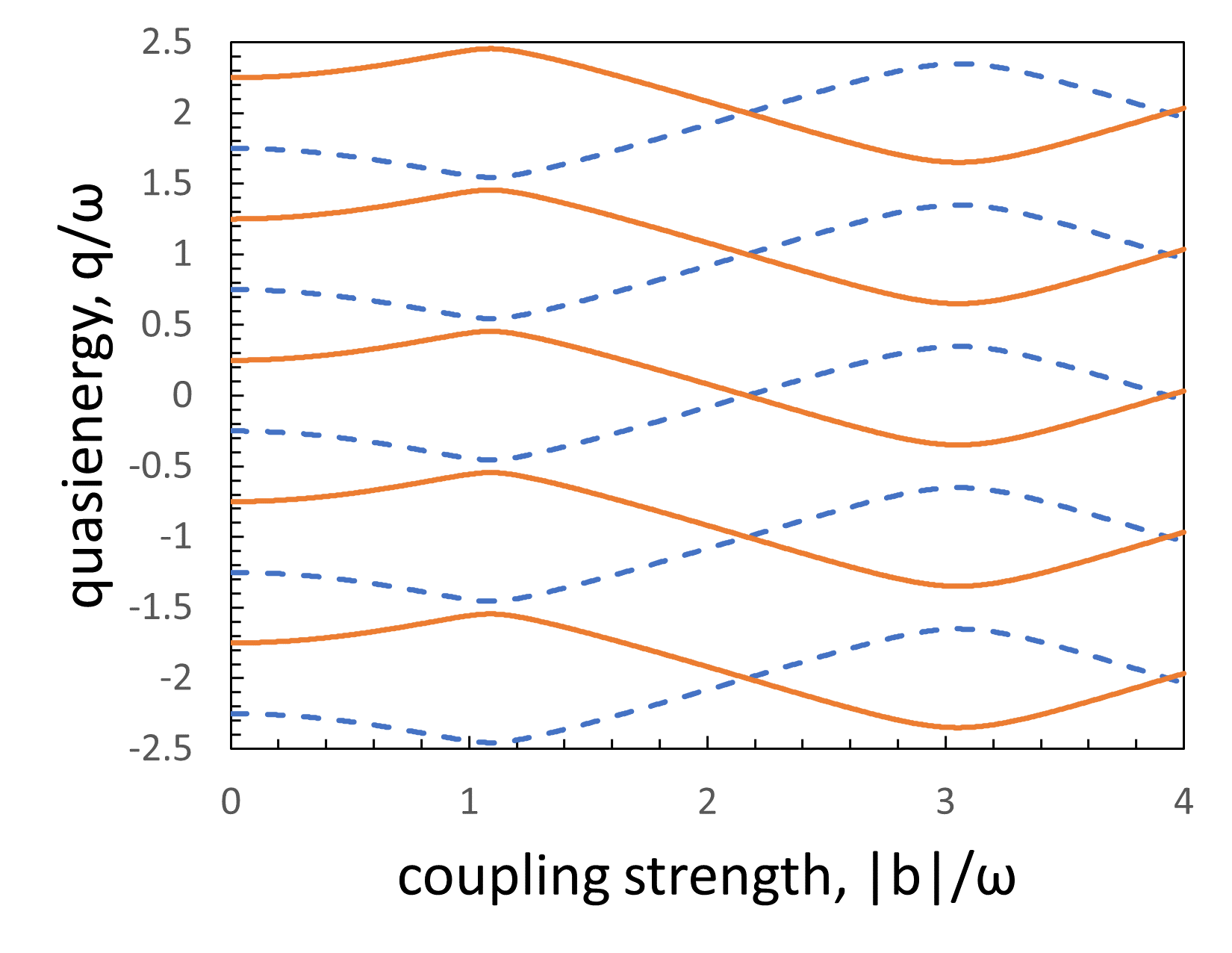}
    \caption{
    Quasienergies of the Hamiltonian (\ref{cosdriven}) with $\omega_0/\omega =2.5$ plotted against coupling strength $b$.
    Solid (dashed) lines show those for Floquet states approaching $|2m\rangle$ ($|1m\rangle$) as $b\rightarrow0$.}
    \label{figspec3}
    \end{center}
\end{figure}

Let us discuss it in the extended Hilbert space.
The basis of the extended Hilbert space can be taken as follows:
\begin{align}
    \label{eq_basis_apdxA}
    | 1 m \rangle =
    \left(
        \begin{array}{c}
            1 \\
            0
        \end{array}
    \right)
    e^{im\omega t},
    | 2 m\rangle =
    \left(
        \begin{array}{c}
            0 \\
            1
        \end{array}
    \right)
    e^{im\omega t},
\end{align}
and the operators of the extended Hilbert space $H_{CF}$ and $P_{T/2,F}$ are expressed in the basis (\ref{eq_basis_apdxA}) as 
\begin{align}
    P_{T/2,F} = 
    \left(
        \begin{array}{cccccccc}
            \ddots & -1 & 0 & 0 & 0 & 0 & 0 & \iddots \\
            \cdots & 0 & 1 & 0 & 0 & 0 & 0 & \cdots \\
            \cdots & 0 & 0 & 1 & 0 & 0 & 0 & \cdots \\
            \cdots & 0 & 0 & 0 & -1 & 0 & 0 & \cdots \\
            \cdots & 0 & 0 & 0 & 0 & -1 & 0 & \cdots \\
            \iddots & 0 & 0 & 0 & 0 & 0 & 1 & \ddots
        \end{array}
    \right),
\end{align}
\begin{align}
    H_{CF} = 
    \left(
        \begin{array}{cccccccc}
            \ddots & \frac{\omega_0}{2} -\omega & 0 & 0 & b & 0 & 0 & \iddots \\
            \cdots & 0 & -\frac{\omega_0}{2} - \omega & b^* & 0 & 0 & 0 & \cdots \\
            \cdots & 0 & b & \frac{\omega_0}{2} & 0 & 0 & b & \cdots \\
            \cdots & b^* & 0 & 0 & -\frac{\omega_0}{2} & b^* & 0 & \cdots \\
            \cdots & 0 & 0 & 0 & b & \frac{\omega_0}{2} + \omega & 0 & \cdots \\
            \iddots & 0 & 0 & b^* & 0 & 0 & -\frac{\omega_0}{2} + \omega & \ddots
        \end{array}
    \right).
\end{align}
This Hamiltonian is infinite dimensional, so we calculate its spectrum numerically, that is given in Fig.~\ref{figspec3}.

\section{Complete set of solvable Hamiltonians for two-level systems}
\label{apdxB}
In this section, we give an explicit expression of Hamiltonians with CDS for two-level systems.
We also see that, for two-level systems, such a Hamiltonian is equivalent to a RWA Hamiltonian with unitary transformations.

Since the generator $A$ is Hermitian it can be diagonalized, and in the base of the Pauli matrix $\{\sigma_i\}_{i=1,2,3}$ it can be written as
\begin{align}
    VAV^\dagger = d_3\sigma_3=\text{diag}(d_3,-d_3),
\end{align}
where $d_3 \in \mathbb{R}$, and $V$ is a unitary matrix.
According to Eq.~(\ref{A=alpha}), $d_3$ must be written as $d_3 = \frac{K}{2}\omega$, where $K \in \mathbb{Z}$.
By substituting this expression to $A$, we find
\begin{align}
    VAV^\dagger = \frac{K\omega}{2} \sigma_3.\label{eq_quantization_apdxB}
\end{align}
Since $K$ can be absorbed at a rational multiple of the reference angular frequency $\omega$, there is no freedom of choice in the eigenvalues of the generator $A$ in the two-level system up to unitary transformation.

One can see that if the generator $A$ is diagonal, the corresponding Hamiltonian has the shape of the Hamiltonian of the Rabi oscillation (\ref{eq_RWhamiltonian}) for any $H_0$.
In addition, since the choice of $H_0$ is arbitrary and any Hermitian matrix can be represented by $H_0-A$, as we discussed in \ref{sec_integrability}, the following statement holds; in the case of a two-level system, for any traceless Hamiltonian $H(t)$ with CDS, there is a combination of parameters $(\omega_0, \omega, b)$ in (\ref{eq_RWhamiltonian}), its spectra match with the spectra of the Hamiltonian of the Rabi oscillation.

Here we solve Eq. (\ref{eq_dH}) directly and show the explicit form of Hamiltonians with the CDS for two-level systems.
In the base of the Pauli matrices the generator $A$ and Hamiltonian $H(t)$ can be expressed as $A=\sum_i a_i \sigma_i$, $H(t)=\sum_i h_i(t) \sigma_i$.
We introduce the following 3-dimensional real-valued vectors:
\begin{align}
    \bm{a}
    =
    \left(
        \begin{array}{c}
            a_1\\
            a_2\\
            a_3
        \end{array}
    \right),
    \bm{h}
    =
    \left(
        \begin{array}{c}
            h_1\\
            h_2\\
            h_3
        \end{array}
    \right).
\end{align}
and Eq. (\ref{eq_dH}) is written as follows:
\begin{align}
    \label{eq_precession}
    \frac{d}{dt} \bm{h} = \bm{a} \times \bm{h}.
\end{align}
Therefore, in the case of a two-level system the vector $\bm{h}(t)$ constructed from the Hamiltonian moves on an orbit of precession around $\bm {a}$ when its degrees of freedom are considered as an orbit of a particle.
Here, we define
\begin{align}
    \tan\theta = \frac{a_1}{a_2}, \tan\phi = \frac{a_3}{\sqrt{a_1^2+a_2^2}}.
\end{align}
By solving Eq. (\ref{eq_precession}), we find the correspooinding Hamiltonican can be written by
\begin{align}
    &H(t) =\nonumber\\
    &\left(
        \begin{array}{cc}
            C \sin\phi \sin\omega t + D\cos\phi & e^{-i\theta}(C(\cos\omega t-i\cos\phi \sin\omega t) -iD\sin\phi)\\
            e^{i\theta}(C\cos\omega t+iD(\cos\phi \sin\omega t +\sin\phi)) & - C \sin\phi \sin\omega t - D\cos\phi
        \end{array}
    \right),
\end{align}
where $C, D$ are arbitrary constants.

Finally, we briefly discuss the extension of the above argument to $N$-level systems. Since the generator $A$ is a traceless Hermitian matrix, it can be written as a linear combination of $N-1$ diagonal (Cartan) generators of SU($N$) group. And thus, the quantization condition corresponding to (\ref{eq_quantization_apdxB}) can also be found to satisfy Eq.~(\ref{A=alpha}). Additionally the time-evolution of Hamiltonian with CDS can be written by using structure constant of the  SU($N$) algebra. Concretely speaking, Eq.~\eqref{eq_precession} for the $N=2$ case can be represented as
\begin{align}
 \frac{d}{dt} h_i = \epsilon_{ijk} a_j h_k,   
\end{align}
where the repeated indices $j$ and $k$ are implicitly summed. This equation can be generalized by replacing $\epsilon_{ijk}$ by the structure constant $f_{ijk}$ for the SU($N$) algebra.

\printbibliography 

@article{Eckardt2017,
  title={Colloquium: Atomic quantum gases in periodically driven optical lattices},
  author={Eckardt, Andr{\'e}},
  journal={Reviews of Modern Physics},
  volume={89},
  number={1},
  pages={011004},
  year={2017},
  publisher={APS}
}

@article{Zhao2019,
  title = {Floquet time spirals and stable discrete-time quasicrystals in quasiperiodically driven quantum many-body systems},
  author = {Zhao, Hongzheng and Mintert, Florian and Knolle, Johannes},
  journal = {Phys. Rev. B},
  volume = {100},
  issue = {13},
  pages = {134302},
  numpages = {7},
  year = {2019},
  month = {Oct},
  publisher = {American Physical Society},
  doi = {10.1103/PhysRevB.100.134302},
  url = {https://link.aps.org/doi/10.1103/PhysRevB.100.134302}
}

@Article{Ikeda2021,
	title={{Nonequilibrium steady states in the Floquet-Lindblad systems: van  Vleck's high-frequency expansion approach}},
	author={Tatsuhiko N. Ikeda and Koki Chinzei and Masahiro Sato},
	journal={SciPost Phys. Core},
	volume={4},
	pages={033},
	year={2021},
	publisher={SciPost},
	doi={10.21468/SciPostPhysCore.4.4.033},
	url={https://scipost.org/10.21468/SciPostPhysCore.4.4.033},
}

@article{Nishimura2022,
  title = {Floquet Engineering Using Pulse Driving in a Diamond Two-Level System Under Large-Amplitude Modulation},
  author = {Nishimura, Shunsuke and Itoh, Kohei M. and Ishi-Hayase, Junko and Sasaki, Kento and Kobayashi, Kensuke},
  journal = {Phys. Rev. Appl.},
  volume = {18},
  issue = {6},
  pages = {064023},
  numpages = {12},
  year = {2022},
  month = {Dec},
  publisher = {American Physical Society},
  doi = {10.1103/PhysRevApplied.18.064023},
  url = {https://link.aps.org/doi/10.1103/PhysRevApplied.18.064023}
}

@article{Rondin2014,
doi = {10.1088/0034-4885/77/5/056503},
url = {https://dx.doi.org/10.1088/0034-4885/77/5/056503},
year = {2014},
month = {may},
publisher = {IOP Publishing},
volume = {77},
number = {5},
pages = {056503},
author = {L Rondin and J-P Tetienne and T Hingant and J-F Roch and P Maletinsky and V Jacques},
title = {Magnetometry with nitrogen-vacancy defects in diamond},
journal = {Reports on Progress in Physics}
}

@article{Floquet1883,
    title = {{Sur les {\'{e}}quations diff{\'{e}}rentielles lin{\'{e}}aires {\`{a}} coefficients p{\'{e}}riodiques}},
    year = {1883},
    journal = {Annales scientifiques de l'{\'{E}}cole Normale Sup{\'{e}}rieure},
    author = {Floquet, G},
    pages = {47--88},
    volume = {2},
    publisher = {Elsevier},
    url = {http://archive.numdam.org/articles/10.24033/asens.220/ https://doi.org/10.24033/asens.220},
    doi = {10.24033/asens.220},
}

@article{Rabi1937,
    title = {{Space Quantization in a Gyrating Magnetic Field}},
    year = {1937},
    journal = {Physical Review},
    author = {Rabi, I I},
    number = {8},
    month = {4},
    pages = {652--654},
    volume = {51},
    publisher = {American Physical Society},
    url = {https://link.aps.org/doi/10.1103/PhysRev.51.652},
    doi = {10.1103/PhysRev.51.652}
}

@article{Oka2019,
    title = {{Floquet Engineering of Quantum Materials}},
    year = {2019},
    journal = {Annual Review of Condensed Matter Physics},
    author = {Oka, Takashi and Kitamura, Sota},
    number = {1},
    month = {3},
    pages = {387--408},
    volume = {10},
    publisher = {Annual Reviews},
    url = {https://doi.org/10.1146/annurev-conmatphys-031218-013423},
    doi = {10.1146/annurev-conmatphys-031218-013423},
    issn = {1947-5454}
}

@article{Holthaus2015,
    title = {{Floquet engineering with quasienergy bands of periodically driven optical lattices}},
    year = {2015},
    journal = {J. Phys. B: At. Mol. Opt. Phys.},
    author = {Holthaus, Martin},
    number = {1},
    pages = {13001},
    volume = {49},
    publisher = {IOP Publishing},
    url = {http://dx.doi.org/10.1088/0953-4075/49/1/013001},
    doi = {10.1088/0953-4075/49/1/013001},
    issn = {0953-4075}
}

@article{Rudner2020,
    title = {{The Floquet Engineer's Handbook}},
    year = {2020},
    journal = {arXiv:2003.08252},
    author = {Rudner, Mark S. and Lindner, Netanel H.},
    month = {3},
    url = {https://arxiv.org/abs/2003.08252v2},
    arxivId = {2003.08252}
}

@article{Bukov2015,
    title = {{Universal high-frequency behavior of periodically driven systems: from dynamical stabilization to Floquet engineering}},
    year = {2015},
    journal = {Advances in Physics},
    author = {Bukov, Marin and D'Alessio, Luca and Polkovnikov, Anatoli},
    number = {2},
    pages = {139--226},
    volume = {64},
    publisher = {Taylor {\&} Francis},
    url = {https://doi.org/10.1080/00018732.2015.1055918},
    doi = {10.1080/00018732.2015.1055918}
}

@article{Mikawa2023,
  title = {Electron-spin double resonance of nitrogen-vacancy centers in diamond under a strong driving field},
  author = {Mikawa, Takumi and Okaniwa, Ryusei and Matsuzaki, Yuichiro and Tokuda, Norio and Ishi-Hayase, Junko},
  journal = {Phys. Rev. A},
  volume = {108},
  issue = {1},
  pages = {012610},
  numpages = {11},
  year = {2023},
  month = {Jul},
  publisher = {American Physical Society},
  doi = {10.1103/PhysRevA.108.012610},
  url = {https://link.aps.org/doi/10.1103/PhysRevA.108.012610}
}

@article{Engelhardt2021,
  title = {Dynamical Symmetries and Symmetry-Protected Selection Rules in Periodically Driven Quantum Systems},
  author = {Engelhardt, Georg and Cao, Jianshu},
  journal = {Phys. Rev. Lett.},
  volume = {126},
  issue = {9},
  pages = {090601},
  numpages = {7},
  year = {2021},
  month = {Mar},
  publisher = {American Physical Society},
  doi = {10.1103/PhysRevLett.126.090601},
  url = {https://link.aps.org/doi/10.1103/PhysRevLett.126.090601}
}

@article{Wang2021,
  title = {Observation of Symmetry-Protected Selection Rules in Periodically Driven Quantum Systems},
  author = {Wang, Guoqing and Li, Changhao and Cappellaro, Paola},
  journal = {Phys. Rev. Lett.},
  volume = {127},
  issue = {14},
  pages = {140604},
  numpages = {7},
  year = {2021},
  month = {Sep},
  publisher = {American Physical Society},
  doi = {10.1103/PhysRevLett.127.140604},
  url = {https://link.aps.org/doi/10.1103/PhysRevLett.127.140604}
}

@article{Shirley1965,
  title = {{Solution of the Schr\"odinger Equation with a Hamiltonian Periodic in Time}},
  author = {Shirley, Jon H.},
  journal = {Phys. Rev.},
  volume = {138},
  issue = {4B},
  pages = {B979--B987},
  numpages = {0},
  year = {1965},
  month = {May},
  publisher = {American Physical Society},
  doi = {10.1103/PhysRev.138.B979},
  url = {https://link.aps.org/doi/10.1103/PhysRev.138.B979}
}

@article{ikeda2020general,
  title={General description for nonequilibrium steady states in periodically driven dissipative quantum systems},
  author={Ikeda, Tatsuhiko N and Sato, Masahiro},
  journal={Science Advances},
  volume={6},
  number={27},
  pages={eabb4019},
  year={2020},
  publisher={American Association for the Advancement of Science},
  url={https://www.science.org/doi/full/10.1126/sciadv.abb4019}
}

@article{alon1998selection,
  title={Selection rules for the high harmonic generation spectra},
  author={Alon, Ofir E and Averbukh, Vitali and Moiseyev, Nimrod},
  journal={Physical review letters},
  volume={80},
  number={17},
  pages={3743},
  year={1998},
  publisher={APS},
  url={https://journals.aps.org/prl/abstract/10.1103/PhysRevLett.80.3743}
}

@article{neufeld2019floquet,
  title={Floquet group theory and its application to selection rules in harmonic generation},
  author={Neufeld, Ofer and Podolsky, Daniel and Cohen, Oren},
  journal={Nature communications},
  volume={10},
  number={1},
  pages={405},
  year={2019},
  publisher={Nature Publishing Group UK London},
  url={https://www.nature.com/articles/s41467-018-07935-y}
}

@article{sambe1973steady,
  title={Steady states and quasienergies of a quantum-mechanical system in an oscillating field},
  author={Sambe, Hideo},
  journal={Physical Review A},
  volume={7},
  number={6},
  pages={2203},
  year={1973},
  publisher={APS},
  url={https://journals.aps.org/pra/abstract/10.1103/PhysRevA.7.2203}
}

@article{ikeda2022floquet,
  title={Floquet-Landau-Zener interferometry: Usefulness of the Floquet theory in pulse-laser-driven systems},
  author={Ikeda, Tatsuhiko N and Tanaka, Satoshi and Kayanuma, Yosuke},
  journal={Physical Review Research},
  volume={4},
  number={3},
  pages={033075},
  year={2022},
  publisher={APS},
  url={https://journals.aps.org/prresearch/abstract/10.1103/PhysRevResearch.4.033075}
}

@article{iadecola2013generalized,
  title={Generalized energy and time-translation invariance in a driven dissipative system},
  author={Iadecola, Thomas and Chamon, Claudio and Jackiw, Roman and Pi, So-Young},
  journal={Physical Review B},
  volume={88},
  number={10},
  pages={104302},
  year={2013},
  publisher={APS},
  url={https://journals.aps.org/prb/abstract/10.1103/PhysRevB.88.104302}
}

@article{pizzi2019period,
  title={Period-n discrete time crystals and quasicrystals with ultracold bosons},
  author={Pizzi, Andrea and Knolle, Johannes and Nunnenkamp, Andreas},
  journal={Physical review letters},
  volume={123},
  number={15},
  pages={150601},
  year={2019},
  publisher={APS},
  url={https://journals.aps.org/prl/abstract/10.1103/PhysRevLett.123.150601}
}

@article{autler1955stark,
  title={Stark effect in rapidly varying fields},
  author={Autler, Stanley H and Townes, Charles H},
  journal={Physical Review},
  volume={100},
  number={2},
  pages={703},
  year={1955},
  publisher={APS},
  url={https://journals.aps.org/pr/abstract/10.1103/PhysRev.100.703}
}

@article{holthaus1994generalized,
  title={Generalized $\pi$ pulses},
  author={Holthaus, Martin and Just, Bettina},
  journal={Physical Review A},
  volume={49},
  number={3},
  pages={1950},
  year={1994},
  publisher={APS},
  url={https://journals.aps.org/pra/abstract/10.1103/PhysRevA.49.1950}
}

@article{alon2002dynamical,
  title={Dynamical symmetries of time-periodic Hamiltonians},
  author={Alon, Ofir E},
  journal={Physical Review A},
  volume={66},
  number={1},
  pages={013414},
  year={2002},
  publisher={APS},
  url={https://journals.aps.org/pra/abstract/10.1103/PhysRevA.66.013414}
}

@article{alon2004atoms,
  title={Atoms, molecules, crystals and nanotubes in laser fields: From dynamical symmetry to selective high-order harmonic generation of soft X-rays},
  author={Alon, Ofir E and Averbukh, Vitali and Moiseyev, Nimrod},
  journal={Advances in Quantum Chemistry},
  volume={47},
  pages={393--421},
  year={2004},
  publisher={Elsevier},
  url={https://www.sciencedirect.com/science/article/abs/pii/S0065327604470221}
}

@article{pisanty2019conservation,
  title={Conservation of torus-knot angular momentum in high-order harmonic generation},
  author={Pisanty, Emilio and Rego, Laura and San Román, Julio and Picón, Antonio and Dorney, Kevin M and Kapteyn, Henry C and Murnane, Margaret M and Plaja, Luis and Lewenstein, Maciej and Hernández-García, Carlos},
  journal={Physical Review Letters},
  volume={122},
  number={20},
  pages={203201},
  year={2019},
  publisher={APS},
  url={https://journals.aps.org/prl/abstract/10.1103/PhysRevLett.122.203201}
}

@article{lerner2023multiscale,
  title={Multiscale dynamical symmetries and selection rules in nonlinear optics},
  author={Lerner, Gavriel and Neufeld, Ofer and Hareli, Liran and Shoulga, Georgiy and Bordo, Eliayu and Fleischer, Avner and Podolsky, Daniel and Bahabad, Alon and Cohen, Oren},
  journal={Science Advances},
  volume={9},
  number={15},
  pages={eade0953},
  year={2023},
  publisher={American Association for the Advancement of Science},
  url={https://www.science.org/doi/full/10.1126/sciadv.ade0953}
}

@article{husimi1953miscellanea,
  title={Miscellanea in elementary quantum mechanics, II},
  author={Husimi, K{\^o}di},
  journal={Progress of Theoretical Physics},
  volume={9},
  number={4},
  pages={381--402},
  year={1953},
  publisher={Oxford University Press},
  url={https://academic.oup.com/ptp/article/9/4/381/1849279?login=false}
}

@book{dittrich1998quantum,
  title={Quantum transport and dissipation},
  author={Dittrich, Thomas and H{\"a}nggi, Peter and Ingold, Gert-Ludwig and Kramer, Bernhard and Sch{\"o}n, Gerd and Zwerger, Wilhelm},
  volume={3},
  year={1998},
  publisher={Wiley-Vch Weinheim}
}

@article{blanes2009magnus,
  title={The Magnus expansion and some of its applications},
  author={Blanes, Sergio and Casas, Fernando and Oteo, Jose-Angel and Ros, Jos{\'e}},
  journal={Physics reports},
  volume={470},
  number={5-6},
  pages={151--238},
  year={2009},
  publisher={Elsevier},
  url={https://www.sciencedirect.com/science/article/abs/pii/S0370157308004092}
}

@article{eckardt2015high,
  title={High-frequency approximation for periodically driven quantum systems from a Floquet-space perspective},
  author={Eckardt, Andr{\'e} and Anisimovas, Egidijus},
  journal={New journal of physics},
  volume={17},
  number={9},
  pages={093039},
  year={2015},
  publisher={IOP Publishing},
  url={https://iopscience.iop.org/article/10.1088/1367-2630/17/9/093039/meta}
}

@article{mikami2016brillouin,
  title={Brillouin-Wigner theory for high-frequency expansion in periodically driven systems: Application to Floquet topological insulators},
  author={Mikami, Takahiro and Kitamura, Sota and Yasuda, Kenji and Tsuji, Naoto and Oka, Takashi and Aoki, Hideo},
  journal={Physical Review B},
  volume={93},
  number={14},
  pages={144307},
  year={2016},
  publisher={APS},
  url={https://journals.aps.org/prb/abstract/10.1103/PhysRevB.93.144307}
}

\end{document}